\documentclass[12pt]{article}
\usepackage{epsfig}

\usepackage{latexsym}
\usepackage{indentfirst}
\usepackage{fancyhdr}
\usepackage{amssymb}
\usepackage{amsmath}
\usepackage{tikzfeynman}   

\usepackage{amsfonts}
\usepackage{pifont}
\usepackage{cite}
\usepackage{color}
\usepackage{slashed}
\usepackage{graphics}
\usepackage{url}
\usepackage{array}
\usepackage{bm}
\usepackage{graphicx}
\usepackage{float}



\usepackage{lieart}
\usepackage{array,multirow}
\usepackage{hhline}
\usepackage{graphicx}
\usepackage{subcaption}

\newcommand{\dr}{{\rm d}}

\begin{document}

\begin{titlepage}
 \begin{flushright}
TTP17-018
 \end{flushright}
   \vskip 1cm
   \begin{center}
    {\Large\bf Dark Matter in $E_6$ Grand Unification}
   
   \vskip 0.2  cm
   \vskip 0.5  cm
Jakob Schwichtenberg$^{\,a,}$\footnote{E-mail: \texttt{jakob.schwichtenberg@kit.edu}},
\\[1mm]
   \vskip 0.7cm
 \end{center}

\centerline{$^{a}$ \it  Institut f\"ur Theoretische Teilchenphysik, 
Karlsruhe Institute of Technology,}
\centerline{\it  Engesserstra{\ss}e 7, D-76128 Karlsruhe, Germany} 
\vspace*{1.5cm}

\begin{abstract}
\noindent
We discuss fermionic dark matter in non-supersymmetric $E_6$ Grand Unification. The fundamental representation of $E_6$ contains, in addition to the standard model fermions, exotic fermions and we argue that one of them is a viable, interesting dark matter candidate. Its stability is guaranteed by a discrete remnant symmetry, which is an unbroken subgroup of the $E_6$ gauge symmetry. We compute the symmetry breaking scales and the effect of possible threshold corrections by solving the renormalization group equations numerically after imposing gauge coupling unification. Since the Yukawa couplings of the exotic and the standard model fermions have a common origin, the mass of the dark matter particles is constrained. We find a mass range of ${10^{8} \text{ GeV} \lesssim m_{DM} \lesssim 10^{12} \text{ GeV}}$ for our $E_6$ dark matter candidate, which is within the reach of next-generation direct detection experiments.
\end{abstract}
\end{titlepage}

\section{Introduction}
\label{sec:intro}

Despite compelling evidence for its existence the nature of dark matter is still unknown \cite{Bertone:2004pz}. 
It is well possible that dark matter (DM) consists of particles which emerge from the breaking chain of some grand unified theory (GUT) and are stable due to a remnant discrete symmetry. The recent revival of non-supersymmetric GUTs \cite{Altarelli:2013lla,Bajc:2005zf,Bertolini:2009qj,Bertolini:2012az,DiLuzio:2011my,Joshipura:2011nn,Buccella:2012kc,Altarelli:2013aqa} motivates us to study dark matter candidates in a non-supersymmetric GUT with gauge group $E_6$. The exceptional rank-$6$ group $E_6$ has one important advantage over the widely studied $SU(5)$ \cite{Georgi:1974sy} and $SO(10)$ \cite{Fritzsch:1974nn} groups when it comes to dark matter: the fundamental representation of $E_6$ contains, in addition to the standard model fermions, several exotic fermions. This means that in a $E_6$ GUT we do not need to add any particles by hand in order to have possible dark matter candidates. 

$E_6$ is popular among GUT model builders \cite{Gursey:1975ki,Shafi:1978gg,Stech:1980fn,Barbieri1980369}, because of attractive features such as the automatic absence of anomalies \cite{Gursey:1975ki} and the fact that all standard model fermions of one generation live in the fundamental representation. In addition, there is one feature that really sets $E_6$ apart from all other popular GUT groups: For example, $SU(5)$ is part of the infinite $SU(N)$ family, $SO(10)$ of the infinite $SO(N)$ family and "describing nature by a group taken from an infinite family does raise an obvious question - why this group and not another?" \cite{Witten:2002ei}. In contrast, there are only five exceptional groups and the only one with non-self conjugate representations, which is necessary to avoid complications with mirror fermions, is $E_6$.

A standard way to ensure the stability of DM is through a discrete symmetry. This discrete symmetry can arise naturally when a gauge symmetry is spontaneously broken \cite{PhysRevLett.62.1221,DeMontigny:1993gy,Batell:2010bp,Petersen:2009ip,Frigerio:2009wf}. The idea of a discrete remnant symmetry has been recently incorporated in $SO(10)$ GUT models \cite{Kadastik:2009cu,PhysRevD.91.095010,Nagata:2015dma}. We show that the stability of the lightest exotic $E_6$ fermion can be guaranteed by a remnant discrete symmetry, which is therefore an ideal dark matter candidate. We start by presenting the particle content of our $E_6$ model and discuss under which conditions the lightest exotic fermion is stable through a remnant symmetry. Afterwards, we analyze the Yukawa sector for all allowed breaking chains and discuss the viability of each exotic fermion as a dark matter candidate.

For the most interesting candidate, the exotic neutrino $N_E$, we discuss an explicit scenario in which $N_E$ is stable and could be detected in the near future. We find that $N_E$ is superheavy $10^{7} \text{ GeV} \lesssim m_{N_E}\lesssim 10^{11} \text{ GeV}$ and therefore similar to superheavy dark matter candidates proposed in earlier studies \cite{Kolb:1998ki}. Such a superheavy dark matter particle can be produced with a correct relic density non-thermally in the early universe.

\section{Particle Content}
The particle content of $E_6$ representations depends on the embedding of the standard model gauge group ${G_{SM} \equiv SU(3)_C \times SU(2)_L \times U(1)_Y}$ in $E_6$. Our standard model embedding is specified in Appendix \ref{sec:smemebdding}. As usual, the gauge bosons live in the adjoint representation, which is $78$-dimensional for $E_6$. The fermions (all taken to be left-handed) are contained in the fundamental $27$-dimensional representation $\Psi$ of $E_6$. The particle content of the fermionic $27$, for our standard model embedding, is best understood by considering the decomposition under $SO(10)$
\begin{equation}
\label{eq:fermionso10decomposition}
\Psi =\Psi_1 \oplus \Psi_{10} \oplus \Psi_{16} \, .
\end{equation}
$\Psi_{16}$ contains the $15$ standard model fermions of one generation plus the charge conjugated right-handed neutrino $\nu_R^c$. The fermions in the $10$ are vector-like, because the $10$ is a self-conjugate $SO(10)$ representation. It contains an exotic down-type quark $D$ plus an exotic lepton doublet $( N_E,E)$. In addition, we have an $SO(10)$ singlet $s$. This is summarized in Table \ref{tab:fermionsin27}.

 \begin{table}
\begin{tabular}{ >{$}l<{$} | >{$}l<{$} |  >{$}l<{$} |  >{$}l<{$} | >{$}l<{$} }
\text{Name}  &SO(10)&  2_L2_R4_C &  3_C 2_L 1_Y  &\text{Weight}  \\ \hline
u_L^{\text{red}} & 16 &(2,1,4) & (3,2, \frac{1}{6}) & \weight{1, 0, -1, 0, 0, 1} \\
 u_L^{\text{blue}} & 16 &(2,1,4) & (3,2, \frac{1}{6}) & \weight{-1, 1, -1, 0, 0, 1} \\
  u_L^{\text{green}} & 16 &(2,1,4) & (3,2, \frac{1}{6}) & \weight{0, -1, 0, 0, 0, 1} \\
 d_L^{\text{red}} & 16 & (2,1,4) & (3,2, \frac{1}{6}) & \weight{1, 0, 0, 0, 0, -1} \\
d_L^{\text{blue}} & 16 &(2,1,4) & (3,2, \frac{1}{6})& \weight{-1, 1, 0, 0, 0, -1} \\
 d_L^{\text{green}} & 16 &(2,1,4) & (3,2, \frac{1}{6}) & \weight{0, -1, 1, 0, 0, -1} \\
 \nu_L & 16  &(2,1,4)& (1,2, -\frac{1}{2}) & \weight{0, 0, 0, -1, 1, 1} \\
 e_L & 16 &(2,1,4) & (1,2, -\frac{1}{2})  & \weight{0, 0, 1, -1, 1, -1} \\
 
 \lbrack u_R^{\text{red}}  \rbrack^c & 16 & (1,2,\overline{4}) & (3,1, -\frac{2}{3})  & \weight{-1, 0, 0, 0, 1, 0} \\
  \lbrack u_R^{\text{blue}} \rbrack^c & 16 & (1,2,\overline{4})& (3,1, -\frac{2}{3})& \weight{1, -1, 0, 0, 1, 0} \\
 \lbrack u_R^{\text{green}}\rbrack^c & 16& (1,2,\overline{4}) & (3,1, -\frac{2}{3}) & \weight{0, 1, -1, 0, 1, 0} \\
\lbrack d_R^{\text{red}}\rbrack^c & 16 & (1,2,\overline{4}) & (3,1, \frac{1}{3}) & \weight{-1, 0, 1, -1, 0, 0} \\
  \lbrack d_R^{\text{blue}}\rbrack^c & 16 & (1,2,\overline{4}) & (3,1, \frac{1}{3}) & \weight{1, -1, 1, -1, 0, 0} \\
 \lbrack d_R^{\text{green}}\rbrack^c & 16  & (1,2,\overline{4})& (3,1, \frac{1}{3}) & \weight{0, 1, 0, -1, 0, 0} \\
  \lbrack \nu_R\rbrack^c & 16 & (1,2,\overline{4}) & (1,2, 0)& \weight{0, 0, -1, 1, 0, 0} \\
   \lbrack e_R\rbrack^c & 16 & (1,2,\overline{4}) & (1,2, 1)& \weight{0, 0, 0, 0, -1, 0} \\
   
   & &  & & \\
   
 D^{\text{red}} & 10 & (1,1, 6) & (3,1, -\frac{1}{3})  & \weight{1, 0, 0, 0, 0, 0} \\
 D^{\text{blue}} & 10  & (1,1, 6) & (3,1, -\frac{1}{3}) & \weight{-1, 1, 0, 0, 0, 0} \\
 D^{\text{green}} & 10  & (1,1, 6)& (3,1, -\frac{1}{3})  & \weight{0, -1, 1, 0, 0, 0} \\
 N_E & 10  & (2,2, 1) & (1,2, -\frac{1}{2}) & \weight{0, 0, -1, 1, 0, 1} \\
 E & 10 & (2,2, 1) & (1,2, -\frac{1}{2}) & \weight{0, 0, 0, 1, 0, -1} \\
 \lbrack D^{\text{red}}\rbrack^c & 10   & (1,1, 6)& (3,1, \frac{1}{3})  & \weight{-1, 0, 0, 1, -1, 0} \\
 \lbrack D^{\text{blue}}\rbrack^c & 10  & (1,1, 6)&  (3,1, \frac{1}{3})  & \weight{1, -1, 0, 1, -1, 0} \\
 \lbrack D^{\text{green}}\rbrack^c & 10 & (1,1, 6) & (3,1, \frac{1}{3})  & \weight{0, 1, -1, 1, -1, 0} \\
 \lbrack N_E\rbrack^c & 10  & (2,2, \frac{1}{2}) & (1,2, 1) & \weight{0, 0, 1, 0, -1, -1} \\
 \lbrack E\rbrack^c & 10 & (2,2, \frac{1}{2}) & (1,2, 1) & \weight{0, 0, 0, 0, -1, 1} \\
 
  & &  & & \\
 
 s & 1 & (1,1,1) & (1,1,0) & \weight{0, 0, 0, -1, 1, 0} \\
\end{tabular}
 \caption{Fermions in the fundamental $27$-dimensional representation of $E_6$ with the corresponding $SO(10)$, Pati-Salam and standard model representations. The superscript $c$ denotes charge conjugation. Our standard model embedding is specified in Appendix \ref{sec:smemebdding}.}
\label{tab:fermionsin27}
\end{table}

\noindent We assume that all the symmetry breaking is solely done by Higgs fields that couple to fermions. The corresponding scalar representations are found from the decomposition \cite{Slansky:1981yr}
\begin{equation}
\label{eq:2727product}  \overline{27} \otimes  \overline{27} =  27_s \oplus 351'_s \oplus 351_a, 
\end{equation}  
where the subscripts $s$ and $a$ denote symmetric and antisymmetric, respectively. 

\section{Stability of the Lightest Exotic Fermion}

\noindent Before we discuss possible breaking chains, we derive some restrictions from our requirement of a stable dark matter candidate among the exotic fermions. With our restriction to Higgs fields that couple to fermions, the only viable first intermediate symmetry in accordance with Michel's conjecture\footnote{Michel's conjecture states that minima of Higgs potentials correspond to vacuum configurations that break a given gauge group to a \textbf{maximal} subgroup. Although it is well known that this conjecture is not universally true \cite{Abud:1984ni}, "it expresses the maximizing tendency very well. Even the counter-examples are only slightly less than maximal" \cite{9780521347853}. } is $SO(10)$  \cite{Michel:1979vv}, because the $27$, $351'$ and $351$ contain no singlet under any other viable\footnote{The $27$ branches under $F_4 \subset E_6$ as $27= 1 \oplus 26$ and it is possible to break $E_6 \rightarrow F_4 $, for example, when the linear combination $\weight{0,1,0,-1,0,0}+\weight{1,-1,0,1,-1,0}+\weight{-1,0,0,0,1,0}$ gets a VEV. Nevertheless, the $26$ is a self-conjugate representation and therefore this breaking requires a standard model embedding such that there isn't enough space in one $27$ for all standard model fermions of one generation. Therefore $F_4$ is not an attractive intermediate symmetry.} maximal subgroup. (This is a necessary but not sufficient condition \cite{9780521347853}).

\noindent When $E_6$ breaks to $SO(10)$, necessarily a $U(1)$ factor gets broken, because the rank of $E_6$ is $6$ and the rank of $SO(10)$ is $5$. Therefore it is possible that the vacuum remains invariant under a discrete remnant symmetry \cite{PhysRevLett.62.1221}. Under $E_6 \supset SO(10)\times U(1)$ we have the decompositions \cite{Slansky:1981yr}
\begin{align}
\label{eq:e6so10decomposition}
27 &= 1_4 \oplus 10_{-2} \oplus 16_1 \\
78 &=  1_{0}  \oplus 16_{-3} \oplus  \overline{16}_{3} \oplus 45_0   \\
351' &= 1_{-8} \oplus 10_{-2}  \oplus \overline{16}_{-5} \oplus 54_4 \oplus \overline{126}_{-2} \oplus 144_{1} \\
351 &=  10_{-2}  \oplus \overline{16}_{-5} \oplus 16_{1} \oplus 45_{4} \oplus 120_{-2} \oplus 144_{1} \, .
\end{align}
This tells us, for example, that a remnant $\mathbb{Z}_8$ symmetry remains when the $SO(10)$ singlet in $351'$ is responsible for the breaking $E_6 \rightarrow SO(10)$.

Furthermore, when one of the Higgses in the $\overline{126}$ gets a nonzero vacuum expectation value (VEV), which is necessary for a superheavy Majorana mass of the right-handed neutrino, we are left with a  $\mathbb{Z}_2$ symmetry. We have, of course, $\mathbb{Z}_2 \subset \mathbb{Z}_8 $ and therefore label each of the representations in Eq.~\eqref{eq:e6so10decomposition} by their $\mathbb{Z}_2$ quantum numbers (denoted with a $+$ and $-$). Concretely, we have under $SO(10) \times \mathbb{Z}_2$: 
\begin{align}\label{eq:higgwithevenz}
 27 &= 1^+ \oplus 10^+ \oplus 16^- \\
 78 &=  1^+  \oplus 16^- \oplus  \overline{16}^+ \oplus 45^+   \\
 351' &= 1^+ \oplus 10^+  \oplus \overline{16}^- \oplus 54^+ \oplus \overline{126}^+ \oplus 144^- \\
 351 &=  10^+  \oplus \overline{16}^- \oplus 16^- \oplus 45^+ \oplus 120^+ \oplus 144^- \,.
\end{align}
This $\mathbb{Z}_2$ symmetry remains an exact symmetry as long as only Higgs fields with even $(+)$ $\mathbb{Z}_2$ charge get a VEV. In order to have a stable dark matter candidate among the exotic fermions, we need a breaking chain
\begin{equation}
E_6 \rightarrow \ldots \rightarrow U(1)_Y \times SU(2)_L \times SU(3)_C \times \mathbb{Z}_2  \, .
\end{equation}
Then the lightest fermion in the reducible $10^+ \oplus 1^+$ representation cannot decay into lighter fermions in the $16^-$ \cite{Frigerio:2009wf}, i.e. into standard model fermions. In other words, then the lightest fermion in the $10^+ \oplus 1^+$ representation is stable. This is only correct if no boson with odd $(-)$ $\mathbb{Z}_2$ charge is lighter than our lightest exotic fermion, which is in accordance with the extended survival hypothesis \cite{DelAguila198160,Mohapatra:1982aq}.\footnote{We discuss the implications of the extended survival hypothesis for our model in section \ref{sec:rgerunnining}.} The stability argument can also be formulated more compactly, by defining a new $Z_2$ symmetry from the remnant symmetry: $Z_2' = (-1)^{2s} Z_2 $, where $s$ denotes spin and $Z_2$ the discrete remnant symmetry. Under this new symmetry, we have for the fermions, $27 = 1^- \oplus 10^- \oplus 16^+$ whereas the boson charges stay the same. Thus, we can now simply say that the lightest particle with odd $Z_2'$ symmetry is stable, which in our case is a fermion.

\section{Breaking Chains}
\label{sec:breakingchains}

\begin{figure}[t]
\includegraphics[width=1.0\textwidth]{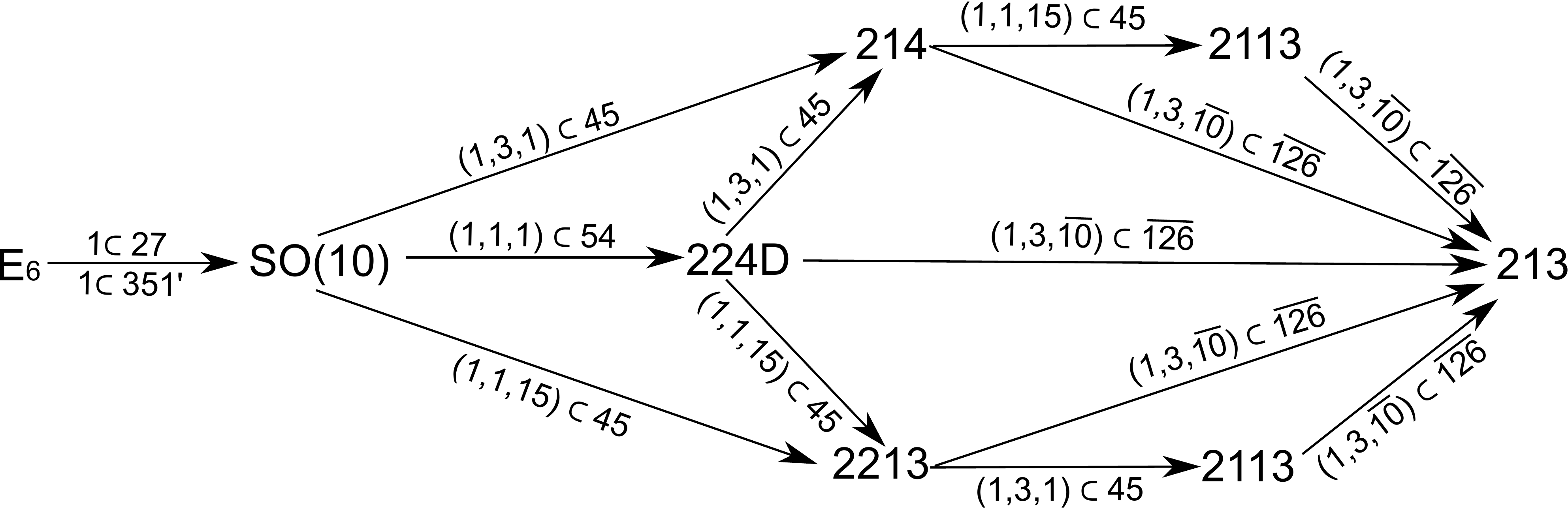} 
\caption{Diagrammatic sketch of possible breaking chains. Here, for example, $213$ denotes $SU(2) \times U(1) \times SU(3)$ and $D$ denotes $D$-parity. The first step in our breaking chain $ E_6 \rightarrow SO(10)$ can be achieved either through the $SO(10)$ singlet in $27$ or through the $SO(10)$ singlet in $351'$. For all further breaking steps the $SO(10)$ representation and the corresponding Pati-Salam submultiplet which are responsible for the breaking are shown.}
\label{fig:so10breakingchains}
\end{figure}

Now, we discuss which breaking chains are possible with the Higgs representations with even $\mathbb{Z}_2$ charge as specified in Eq.~\eqref{eq:higgwithevenz}. It is well known that the Standard Model gauge couplings do not unify \cite{Amaldi:1991cn}. However, if there is an intermediate symmetry between $G_{SM}$ and the GUT symmetry, unification is possible. In order to have an intermediate symmetry between $SO(10)$ and $G_{SM}$ that helps with gauge unification, we need to break $SO(10)$ to a subgroup with equal rank. This is necessary, because $SO(10)$ has rank $5$, the standard model gauge group rank $4$ and there is no viable rank $4$ group that helps with gauge unification. The only two representations in Eq.~\eqref{eq:higgwithevenz} with even $\mathbb{Z}_2$ charge that can achieve such a breaking of $SO(10)$ are $54 \subset 351'$ and $45\subset 351$. The possible intermediate symmetries and breaking chains are shown in Figure~\ref{fig:so10breakingchains}. In the next section, we discuss the implications of the various breaking chains for the masses of the exotic fermions.

\section{Yukawa Sector}
\label{sec:exoticyukawas}

The Yukawa sector above the $E_6$ scale reads \cite{Stech:2003sb}
\begin{equation}
\label{eq:yukawalagrangiane6}
\mathcal{L}_{\text{Y}}  = \Psi^T i \sigma_2 \Psi (Y_{27} \varphi + Y_{351'} \phi + Y_{351} \xi )  + h.c. \, ,
\end{equation}
where $\Psi$ denotes the fermionic $27$, $Y_{i}$ Yukawa couplings and  $\varphi$, $\phi$ and $\xi$ the Higgs representations $27$, $351'$, $351$, respectively. For the $SO(10)$ embedding specified in Appendix \ref{sec:smemebdding}, the Higgs fields that can achieve the breaking $E_6 \rightarrow SO(10)$ are
\begin{align} \label{eq:so10vevs}
\weight{0, 0, 0, -1, 1, 0}  \ &\in \ 27 \notag \\
\weight{0, 0, 0, 2, -2, 0}  \ &\in \ 351' .
\end{align}
A VEV for the $SO(10)$ singlet in the $351'$ yields a mass for the fermionic $SO(10)$ singlet $s$, whereas a VEV for the $SO(10)$ singlet in the $27$ yields a mass for the fermions in the $10 \subset 27$. 

\vspace{0.3cm}

\noindent From Figure \ref{fig:so10breakingchains} we can see that the Pati-Salam submultiplets involved in the further breaking are $(1,1,1) \subset 54$, $(1,3,1) \subset 45$,  $(1,1,15) \subset 45$ and $(1,3,\overline{10}) \subset \overline{126}$. These VEVs yield the following mass terms for the exotic fermions:

\begin{itemize}
\item The VEV in the $54$ that breaks $SO(10)\rightarrow 224D$ reads 
\begin{equation} \label{eq:so10patisalamvev}
   \langle \phi_{54} \rangle  = \sqrt{\frac{5}{12}} \text{diag} \left( \frac{2}{5},\frac{2}{5},\frac{2}{5},\frac{2}{5},\frac{2}{5},\frac{2}{5},\frac{-3}{5},\frac{-3}{5},\frac{-3}{5},\frac{-3}{5} \right)  v_{54} \,.
\end{equation}
We have $10 \otimes 10= 1_s \oplus 54_s \oplus 45_a$ and therefore this VEV yields mass terms for the exotic fermions in the $10$, with $m_D = \frac{2}{3} m_L$, where $L$ denotes the exotic lepton doublet $(N_E,E)$. 
\item The VEV for the Standard Model singlet in $(1,3,1) \subset 45$ yields a mass for the exotic lepton doublet $(N_E,E)$.
\item The VEV for the Standard Model singlet in $(1,1,15) \subset 45$ yields a mass for the exotic quark $D$.
\item The VEV for the Standard Model singlet in $(1,3,\overline{10}) \subset \overline{126}$ yields a mass for the right-handed neutrino $\nu_R$.
\end{itemize}
\vspace{0.3cm}

\noindent With this information at hand, we can now discuss the viability of the various exotic fermions as dark matter candidates.

\section{Candidates}

\subsection[The Exotic Quark $D$]{The Exotic Quark \boldmath{$D$}}
\label{sec:exquark}

The exotic down-type quark $D$ carries hypercharge and color-charge, but no weak-isospin. There are breaking chains where $D$ is the lightest exotic fermion and therefore stable. In such scenarios dark matter would be bound states involving the lightest exotic quark $D$. 

However there are strong bounds on strongly interacting dark matter from direct detection experiments. Although the main goal of direct detection experiments is to detect WIMPs, they are, of course, also sensitive to strongly-interacting dark matter. Such particles would interact several times in the detector and this, together with a local dark matter density of approximately $0.3 \mathrm{ \ \frac{GeV}{cm^3}}$, can be used to exclude large regions of the parameter space for masses below $10^{15}$ GeV \cite{Albuquerque:2003ei}. The remaining regions in the parameter space for strongly-interacting dark matter with a mass below $10^{15}$ GeV are ruled out by the IceCube experiment \cite{Albuquerque:2010bt}.

The Yukawa couplings of the exotic fermions and the Standard Model fermions have a common origin above the $E_6$ scale. Therefore, we expect for the lightest generation of the exotic fermions Yukawa couplings comparable to the Yukawa couplings of the lightest generation of the Standard Model fermions. This, together with an $E_6$ scale below the Planck scale, yields a $D$ mass below $10^{15}$ GeV. Therefore, in the class of $E_6$ models that we consider here, all breaking chains where the exotic quark $D$ is the lightest exotic fermion are ruled out already through direct detection experiments and the IceCube experiment.

\subsection[The Exotic $SO(10)$ Singlet $s$]{The Exotic \boldmath{$SO(10)$} Singlet \boldmath{$s$}}
\label{sec:exsinglet}
From the discussion in Section \ref{sec:exoticyukawas}, we know the there are two possibilities how the exotic singlet $s$ can get a mass: Through a VEV that breaks $E_6$ or  through a VEV that breaks $SU(2)_L \times U(1)_Y$. 

In the first case, if additionally all other exotic fermions are heavier\footnote{This, of course, requires that both $SO(10)$ singlets $1 \subset 27$ and $1  \subset 351'$ develop an $E_6$ scale VEV. Then, depending on the Yukawa couplings $Y_{27}$ and $Y_{351'}$ and the relative magnitude of the VEVs, it is possible that $s$ is the lightest exotic fermion.}, we have a viable but phenomenological rather uninteresting situation. In this scenario, dark matter is a superheavy ($M_s \gtrsim
 10^{10}$ GeV) standard model singlet and it is hard to imagine how such a candidate could ever be detected in experiments.

In the second case, assuming the $E_6$ singlet in the $351'$ does not get a nonzero VEV, $s$ gets only a mass through the breaking of $G_{SM}$. In this scenario, the $E_6$ symmetry is broken through the VEV in the $27$ that yields a superheavy mass for all other exotic fermions. After the breaking of $G_{SM}$, the exotic singlet mixes with the exotic, superheavy neutrino $N_E$. Therefore, in the subspace\footnote{There is no mixing with the other neutral fermions, because otherwise the remnant $\mathbb{Z}_2$ symmetry is broken.} $(s,N_E,N_E^c)$, the mass matrix reads

\begin{equation}
M= \begin{pmatrix}
0 & m & m \\
m & 0 & \mu \\
m & \mu & 0
\end{pmatrix} \, ,
\end{equation}
\begin{sloppypar} 
\noindent where $\mu$ denotes the superheavy mass of $N_E$ and $m$ is an electroweak scale mass. The eigenvalues of this mass matrix are $m_1=-\mu$, $m_2=\frac{1}{2}\left(\mu+\sqrt{8m^2+\mu^2} \right)\approx \mu$ and ${m_3=\frac{1}{2}\left(\mu-\sqrt{8m^2+\mu^2} \right)\approx \frac{-2m^2}{\mu}}$. Proton decay experiments yield a lower bound for the $E_6$ scale of approximately $10^{15.6}$. This means directly that in this scenario the mass of $s$ is far below the Tremaine-Gunn bound \cite{PhysRevLett.42.407}, which is a lower-limit $m>\mathcal{O}($keV) on fermionic dark matter from phase-space arguments. Therefore in this scenario $s$ is superlight and stable, but can not be a dominant component of the observed dark matter density. 
\end{sloppypar}

\subsection[The Exotic Neutrino $N_E$]{The Exotic Neutrino \boldmath{$N_E$}}
\label{sec:exoticneutrinon}
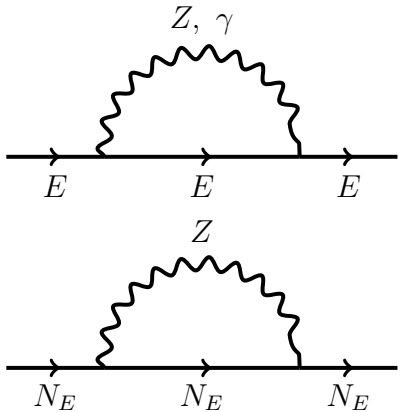
\begin{figure}[t]
\begin{center}
\begin{tikzpicture}[line width=1.5 pt, scale=1.3]
    \draw[fermion] (0,0)--(1,0);
    \draw[fermion] (1,0)--(3,0);
    \draw[fermion] (3,0)--(4,0);
    \draw[vector] (1,0) arc (180:0:1);
    \node at (2, 1.4) {$Z, \ \gamma$};
	\node at (0.5, -0.3) {$E$};
	\node at (2, -0.3) {$E$};
	\node at (3.5, -0.3) {$E$};
\end{tikzpicture}

\begin{tikzpicture}[line width=1.5 pt, scale=1.3]
    \draw[fermion] (0,0)--(1,0);
    \draw[fermion] (1,0)--(3,0);
    \draw[fermion] (3,0)--(4,0);
    \draw[vector] (1,0) arc (180:0:1);
    \node at (2, 1.4) {$Z$};
	\node at (0.5, -0.3) {$N_E$};
	\node at (2, -0.3) {$N_E$};
	\node at (3.5, -0.3) {$N_E$};
\end{tikzpicture}
\end{center}
\caption{Radiative correction that contribute to the mass splitting of a vector-like lepton doublet $(N_E,E)$, with degenerate mass $M$ at tree level.}
\label{fig:radiativecorrectionsexoticfermions}
\end{figure}

In all breaking chains, the exotic electron $E$ and the exotic neutrino $N_E$ have exactly the same tree level mass $M$ before the breaking of $SU(2)_L \times U(1)_Y$. This degeneracy is removed by the radiative corrections shown in Figure \ref{fig:radiativecorrectionsexoticfermions}. The mass splitting can be calculated to be \cite{Sher:1995tc}

\begin{equation} \label{eq:masssplitting}
\Delta M \equiv M_E - M_{N_E} = \frac{\alpha M}{2 \pi} \int_0^1 dx (1+x) \Big[\ln\Big(x^2+\frac{m_Z^2}{M^2}(1-x)\Big)-\ln(x^2)\Big] \, ,
\end{equation}
which vanishes in the limit of an unbroken $SU(2)_L$ symmetry $m_Z \rightarrow 0$. This splitting is extremely insensitive to the tree-level mass $M$ and we have $\Delta M \approx \mathcal{O}$($100$ MeV). Therefore the electrically charged $E$ is always heavier than the neutral $N_E$. 

\vspace{0.3cm}

An interesting aspect of $N_E$ as possible dark matter candidate is that it carries hypercharge. Direct detection experiments are capable of detecting hypercharged dark matter with masses as high as $10^{10}$ GeV \cite{Feldstein:2013uha}. The spin-independent cross section for the interaction of a fermionic dark matter particle with hypercharge $\frac{1}{2}$ with a nucleus via $Z$-boson exchange is given by
\begin{equation}
\label{eq:directdetectioncrosssection}
\sigma_{\chi N} = \frac{G_F^2 \mu_N^2}{2\pi} \frac{1}{4} (N - (1-4\sin^2\theta_W)Z)^2 ,
\end{equation}
where $G_F$ denotes the Fermi constant, $\mu_N$ the reduced mass of the nucleus and dark matter particle, $N$ and $Z$ are the the number of neutrons and protons in the target nucleus and $\theta_W$ denotes the Weinberg angle. The latest exclusion limits from the LUX experiment \cite{Akerib:2015rjg} yield a lower bound of $10^{7.8}$ GeV for hypercharged dark matter. A future experiment like DARWIN \cite{Baudis:2014naa} will be able to detect hypercharged dark matter with a mass as high as $10^{10}$ GeV. Therefore, scenarios where $N_E$ is the lightest exotic fermion with a mass in the range $10^{7.8} \text{\ GeV} < N_E < 10^{10} \text{\ GeV}$ are particularly interesting. In the following, we discuss one such scenario.

\subsubsection{Breaking Chain and Mass Terms}

An attractive breaking chain, where it is possible that the lightest exotic particle is the neutral component of the exotic lepton doublet is

\begin{equation}
    E_6 \to SO(10) \to  SU(2)_L \otimes SU(2)_R \otimes SU(4)_C  \otimes D \to U(1)_Y  \otimes SU(2)_L \otimes  SU(3)_C \, .
    \label{eq:breakingchain}
\end{equation}

As discussed in section \ref{sec:exoticyukawas}, there is an $SO(10)$ singlet in the scalar $27$ representation and another one in the scalar $351'$ representation. Both could be responsible for the breaking of $E_6 \to SO(10)$. The $SO(10)$ singlet in the $351'$ yields mass terms for the exotic singlets $s$ and the $SO(10)$ singlet in the $27$ mass terms for all other exotic particles. Furthermore, the VEV (Eq. \eqref{eq:so10patisalamvev}) that breaks ${SO(10) \to  SU(2)_L \otimes SU(2)_R \otimes SU(4)_C  \otimes D}$ yields additional contributions to the masses of the exotic quarks and exotic leptons. If this VEV would be solely responsible for the masses of the exotics, we would have the mass ratio $m_D=\frac{2}{3} m_L$ at the $SO(10)$ scale, where $m_D$ denotes the mass matrix of the exotic quarks and $m_L$ the mass matrix of the exotic leptons. Therefore, without the additional contribution from the $SO(10)$ singlet VEV in the $27$, the exotic quarks would be lighter than the exotic leptons. 

One could argue that the relation $m_D = \frac{2}{3} m_L$ only holds at the $SO(10)$ scale and could be altered dramatically by the RGE running of the Yukawa couplings. Something similar happens in $SU(5)$ GUTs, where the GUT scale relation $m_b = m_\tau$, becomes at the electroweak scale $m_b \approx 3 m_\tau$ \cite{Buras:1977yy}. 
However, these fermions are much lighter than our exotic fermions and the majority of this mass difference is a result of the running close to the electroweak scale where the gauge couplings are sufficiently distinct.
We can check that for the exotic fermions the effect of the RGE running is too small to reverse the situation, i.e. to yield $m_L < m_D$. For the lightest generation of the exotics the dominant terms in the Yukawa RGEs are proportional to the gauge couplings
\begin{align} \label{eq:yukrgeexotics}
16\pi^2  \frac{\dr Y^{(L_E)}_{351'}}{\dr \ln(u)} &= 
   -\left(\frac{9}{2}g_{2L}^{2}+\frac{9}{2}g_{2R}^{2}
\right) Y^{(L_E)}_{351'} + \ldots\\
16\pi^2  \frac{\dr Y^{(D)}_{351'}}{\dr \ln(u)} &= -\left(15g_{4C}^{2}\right) Y^{(D)}_{351'} + \ldots \, .
\end{align}
The gauge couplings are unified at the $SO(10)$ scale and therefore are not very different at scales a few orders of magnitude below the $SO(10)$ scale. In this breaking chain both $D$ and $L_E$ get a mass at the $SO(10)$ scale ($\approx 10^{16}$ GeV) and therefore a rough estimate for the mass of the lightest generation is $10^{10}$ GeV. The running from the $SO(10)$ scale to $10^{10}$ GeV is not enough to yield $m_L < m_D$. 
Therefore, we conclude that in order to get a scenario where $N_E$ is the lightest exotic fermion and we have the breaking chain as given in Eq. \eqref{eq:breakingchain}, the mass matrices are
\begin{align}
\label{eq:exoticmasses}
m_{s} &=  Y_{351'} \langle \Phi_1 \rangle \, , \\
m_{D} &= Y_{27}  \langle \varphi_1 \rangle +  \frac{1}{\sqrt{15}}   Y^{(D)}_{351'} v_{54} \, ,  \\
m_{(N_E,E)} &=  Y_{27} \langle \varphi_1 \rangle  - \frac{\sqrt{3}}{2\sqrt{5}}     Y^{(L_E)}_{351'}  v_{54} \, .
\end{align}
We can already see here that all exotic fermions are superheavy. This in accordance with the \textit{survival hypothesis} \cite{Georgi:1979md,Barbieri:1979ag}, which states that the only fermions that remain massless before electroweak symmetry breaking, are those that cannot get a mass term which is invariant under the standard model gauge group. All exotic fermions are vector-like and thus can get standard model invariant mass terms. 

\subsubsection{Estimation of the masses and direct detection}

\begin{figure}[t]
\centering
\includegraphics[width=1.0\textwidth]{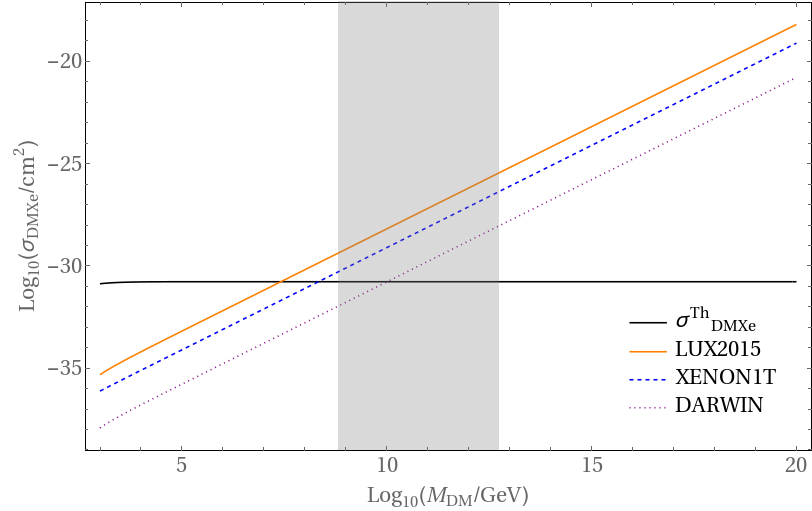}
\caption{Exclusion limit at 90\% C.L from the LUX  experiment \cite{Akerib:2016vxi} and projections for the XENON1T \cite{Aprile:2012zx} and DARWIN \cite{Baudis:2014naa} experiment. The black solid line is the cross section/nucleus for hypercharged dark matter (Eq.~\eqref{eq:directdetectioncrosssection}) and the shaded area indicates our estimate for the mass of the $E_6$ dark matter candidate $N_E$ (Eq. \eqref{eq:massrangeestimate}).}
\label{fig:dmdirectdetection}
\end{figure}

To get an estimate the masses of the exotic fermions we need two things: the Yukawa couplings and the VEVs. We can get a realistic estimate of the Yukawa couplings, by observing that the Yukawa couplings of the standard model particles and of the exotic fermions have a common origin (c.f. Eq. \eqref{eq:yukawalagrangiane6}). After the breaking of the $E_6$ symmetry, the Yukawa sector reads

\begin{align}
\mathcal{L}_{\text{Y}} &=  Y_{27} \Psi_{16}^T i \sigma_2 \Psi_{16} \varphi_{10} + Y_{351'} \Psi_{16}^T i \sigma_2 \Psi_{16} \Phi_{\overline{126}} + Y_{351'} \Psi_{10}^T i \sigma_2 \Psi_{10}  \langle \Phi_{54} \rangle    \\
& \quad +  Y_{351'}  \Psi_1^T i \sigma_2 \Psi_1   \langle \Phi_1 \rangle   + h.c. \, ,
\end{align}
where we neglected all scalar subrepresentations that do not develop a nonzero VEV in our scenario. We can see that the resulting Yukawa sector for the standard model fermions is exactly the same as in $SO(10)$ models with scalars in the $10 \oplus \overline{126}$ representation. The corresponding Yukawa couplings can be fitted such that the standard model fermion observables are correctly reproduced \cite{Dueck:2013gca}. By assuming that the running of the Yukawa couplings between the $E_6$ scale and the $SO(10)$ scale is negligible, we can use these fitted Yukawa couplings to estimate the masses of the exotic fermions. The assumption that the RGE running is here negligible, is reasonable, because there are at most three order of magnitude between the GUT scale and the Planck scale and in addition, only one unified gauge coupling. 
A fit of the Yukawa couplings in an $SO(10)$ model with scalars in the $10 \oplus \overline{126}$ and breaking chain
\begin{equation}
SO(10) \to  SU(2)_L \otimes SU(2)_R \otimes SU(4)_C  \otimes D \to U(1)_Y  \otimes SU(2)_L \otimes  SU(3)_C 
\end{equation}
that takes into account the modified RGEs\footnote{These RGEs are also valid in our model, because the exotic fermions do not mix with the standard model fermions.} between the Pati-Salam and the GUT scale was recently done in Ref. \cite{Meloni:2016rnt}. To estimate the masses of the exotic fermion, we use the best fit-point of this study:
\begin{align}
 Y_{27} &\simeq { \left( {\begin{array}{ccc}
2.21\cdot 10^{-6} & 0 & 0 \\
0 & -1.65\cdot 10^{-3} & 0 \\
0 & 0 & -0.508\end{array} } \right)}\,,\label{eq:h_min}\\
Y_{351'}&\simeq { \left( { \begin{array}{ccc}
3.99\cdot 10^{-6}-2.31\cdot 10^{-5}{\rm i} &5.74\cdot 10^{-6}+1.32\cdot 10^{-4}{\rm i} & -1.55\cdot 10^{-2}-4.10\cdot 10^{-2}{\rm i}\\
5.74\cdot 10^{-6}+1.32\cdot 10^{-4}{\rm i}  & 8.08\cdot 10^{-7}+4.59\cdot 10^{-4}{\rm i}& -0.154+6.25\cdot 10^{-5}{\rm i} \\
 -1.55\cdot 10^{-2}-4.10\cdot 10^{-2}{\rm i}& -0.154+6.25\cdot 10^{-5}{\rm i} & -6.89\cdot 10^{-2}-7.58\cdot 10^{-5}{\rm i}\end{array} } \right)}\, \label{eq:bestfit} \, ,
\end{align}
which was done using $M_{SO(10)} \simeq 1.7 \cdot 10^{15}$ GeV, $M_{PS} \simeq 1.5 \cdot 10^{12}$ GeV and $\alpha_{SO(10)} \simeq 0.027$. These values were computed by using the RGEs for the gauge couplings and the corresponding proton lifetime $\tau_P \approx \frac{M_{SO10}^4 }{m_p^5 \alpha_{SO10}^2}$ is well below the present bound $\tau_P \gtrsim  1.6 \cdot 10^{34}$ yrs for the dominant decay mode $p \rightarrow \pi^0 e^+ $ in non-supersymmetric GUTs \cite{Miura:2016krn}. However, we will show in section \ref{sec:thresholdeffects} that the proton lifetime can be long enough through threshold effects. Therefore, we use in the following instead of $M_{SO(10)} \simeq 1.7 \cdot 10^{15}$ GeV a slightly higher $SO(10)$ scale that is compatible with the current proton decay bounds. 
As noted above, usually we can compute the scales in a GUT model by using the RGEs for the gauge couplings. However, we can not compute the $E_6$ scale, because the couplings are already unified at the $SO(10)$ scale and therefore there is no boundary condition left. In addition, as we will discuss below, the solution of the RGEs for the gauge couplings depends on the masses of the exotic fermions. Therefore, as a first step, we estimate the mass range for the exotic fermions by using Eq. \eqref{eq:bestfit}, Eq. \eqref{eq:exoticmasses}, $M_{SO(10)} \simeq 5 \cdot 10^{15}$ and an $E_6$ VEV between $ 5 \cdot 10^{15}$ GeV and the Planck scale. For the mass of the lightest exotic neutrino, we find
\begin{equation}
6 \cdot 10^8  \text{ GeV} \lesssim m_{N_E} \lesssim  5 \cdot 10^{12} \text{ GeV} \, .
\label{eq:massrangeestimate}
\end{equation}
With this information at hand, we now solve the RGEs for the gauge couplings in a specific scenario.
For\footnote{Here we used $v_{54} \simeq M_{SO(10)}/ \sqrt{4\pi \alpha} $} 
\begin{align}
v_{54} &= -8.5 \cdot 10^{15} \text{ GeV} \notag \\
\langle \varphi_1 \rangle &=8.61 \cdot 10^{16} + 3.24 \cdot 10^{15} i \text{ GeV}
\end{align}
the masses of the exotic quarks and leptons are 
\begin{align}
\{M_L^3,M_L^2,M_L^1 \} &= \{1.53\cdot10^{18} \text{ GeV}, 9.48 \cdot 10^{17} \text{ GeV}, 2.28\cdot 10^{9} \text{ GeV}\} \notag \\
\{M_D^3,M_D^2,M_D^1 \} &= \{1.01\cdot10^{18} \text{ GeV}, 5.59\cdot10^{17} \text{ GeV}, 3.64\cdot 10^{11} \text{ GeV}\} \, .
\label{eq:specificscenario}
\end{align}
Therefore, in this scenario there is only one generation of the exotic fermions present at scales below the GUT scale. In the next section, we investigate the influence of the exotic fermions on the running of the gauge couplings.
As already noted $N_E$ carries hypercharge and could therefore be detected in direct detection experiments. The implication of the mass range in Eq.~\eqref{eq:massrangeestimate} is shown in Figure \ref{fig:dmdirectdetection}.

\subsubsection{RGE running of the gauge couplings}
\label{sec:rgerunnining}

The two-loop RGEs for the gauge couplings are
\begin{equation} \label{eq:twolooprge}
\frac{d\omega_i(\mu)}{d \ln \mu} = - \frac{a_i}{2 \pi} - \sum_j \frac{b_{ij}}{8  \pi^2 \omega_j}, 
\end{equation}
where the indices $i,j$ denote the various subgroups at energy scale $\mu$ and
\begin{equation}
\omega_i = \alpha_i^{-1} = \frac{4\pi}{g_i^2}.
\end{equation}
The coefficients $a_i$ and $b_{ij}$ depend on what particles are present at a given energy scale.
For the scalar masses we invoke the \textit{extended survival hypothesis}, which states that "Higgses acquire the maximum mass compatible with the pattern of symmetry breaking." \cite{delAguila:1980at}. Another point of view is that this a hypothesis of minimal fine tuning \cite{Mohapatra:1982aq}, because only those Higgs fields are light that need to be for the symmetry breaking. The Higgs masses found using this hypothesis are listed in table \ref{tab:higgsmasses}. Take note that for some representations the mass scale is not entirely fixed, but small changes to the spectrum will not change our results dramatically. 
The coefficients for the standard model RGEs are
 \cite{Deshpande:1992au}
\begin{align}
a_{\text{SM}}=\left(
\begin{array}{c}
 \frac{41}{10} , -\frac{19}{6} , -7 
\end{array}
\right) \qquad , \qquad
b_{\text{SM}}= \left(
\begin{array}{ccc}
 \frac{199}{50} & \frac{27}{10} & \frac{44}{5} \\
 \frac{9}{10} & \frac{35}{6} & 12 \\
 \frac{11}{10} & \frac{9}{2} & -26 \\
\end{array}
\right) \, .
\label{eq:smbetacoefficients}
\end{align}
These are valid up to $2.28\cdot 10^{9}$ GeV, where the additional lepton doublet must be taken into account. For simplicity, we use one-loop RGEs above this scale. Using the formulas in Ref. \cite{Jones:1981we} we compute
\begin{equation}
a_{\text{SM+L}}=\left(
\begin{array}{c}
 \frac{43}{10} , -\frac{5}{2} , -7 
\end{array}
\right) \, .\end{equation}
In addition, above $3.64\cdot 10^{11}$ GeV the coefficients change again because of the additional quark. 
\begin{equation}
a_{\text{SM+L+D}}=\left(
\begin{array}{c}
 \frac{79}{18} , -\frac{5}{2} , -\frac{19}{3} 
\end{array}
\right) \, .\end{equation}
Finally, above the Pati-Salam scale we must take additional scalars and gauge bosons into account
\begin{equation}
a_{\text{PS}} = \left(
\begin{array}{c}
 \frac{28}{3} , \frac{28}{3} , 2
\end{array}
\right) \, .
\label{eq:psbetacoefficients}
\end{equation}

\begin{table}[]
\centering
\begin{tabular}{lllll}
 $E_6$  & $SO(10)$  & $2_L2_R4_C$ & $1_Y2_L3_C$ & Mass Scale \\   \hhline{|=====|}
 \multirow{5}{*}{$\varphi (27)$} & $\varphi_{16}$  &  & &    $M_{E_6}$ \\ \cline{2-5}
  & \multirow{3}{*}{$\varphi_{10}$} & $\varphi_{10}^1$(1,1,6) &  & $M_{SO(10)}$  \\ \cline{3-5}
 &  & \multirow{2}{*}{$\varphi_{10}^2$(2,2,1)}  & $\varphi_{10}^{21} (\frac{-1}{2},2,1)$ & $M_{SM}$  \\
 & &   & $\varphi_{10}^{22}$($\frac{1}{2}$,2,1) & $M_{PS}$ \\  \cline{2-5}
 &  $\varphi_{1}$&  &  &   $M_{E_6}$ \\
 \hline
 \multirow{29}{*}{$\Phi (351')$} &  $\Phi_{144}$ & &  & $M_{E_6}$ \\ \cline{2-5}
 & \multirow{19}{*}{$\Phi_{126}$} &  $\Phi_{126}^1 (1,1,6)$   &   & $M_{SO(10)}$  \\  \cline{3-5}
 &  & \multirow{3}{*}{$\Phi_{126}^2 (3,1,10)$ }   & $\Phi_{126}^{21} (-1,3,1) $ & $M_{PS}$ \\
 &&&$\Phi_{126}^{22} (\frac{-1}{3},3,3) $ & $M_{PS}$ \\ 
  &&&$\Phi_{126}^{23} (\frac{1}{3},3,6)$& $M_{PS}$ \\  \cline{3-5}
  &  & \multirow{9}{*}{$\Phi_{126}^3 (1,3,\overline{10})$ }   & *$\Phi_{126}^{31}(0,1,1) $ & $M_{PS}$ \\
&&&*$\Phi_{126}^{32} (1,1,1) $ & $M_{PS}$ \\ 
  &&&$\Phi_{126}^{33} (2,1,1)$& $M_{PS}$ \\ 
   &&&$\Phi_{126}^{34} (\frac{4}{3},1,\overline{3})$& $M_{PS}$ \\ 
    &&&$\Phi_{126}^{35} (\frac{1}{3},1,\overline{3})$& $M_{PS}$ \\ 
     &&&*$\Phi_{126}^{36} (\frac{-2}{3},1,\overline{3})$& $M_{PS}$ \\ 
      &&&$\Phi_{126}^{37} (\frac{-4}{3},1,\overline{6})$& $M_{PS}$ \\ 
       &&&$\Phi_{126}^{38} (\frac{-1}{3},1,\overline{6})$& $M_{PS}$ \\ 
        &&&$\Phi_{126}^{39} (\frac{2}{3},1,\overline{6})$& $M_{PS}$ \\   \cline{3-5}
&  & \multirow{8}{*}{$\Phi_{126}^4 (2,2,15)$ }   & $\Phi_{126}^{41} (\frac{-1}{2},2,1) $ & $M_{PS}$ \\
&&&$\Phi_{126}^{42}(\frac{1}{2},2,1) $ & $M_{PS}$ \\ 
  &&&$\Phi_{126}^{43} (\frac{7}{6},2,3)$& $M_{PS}$ \\ 
   &&&$\Phi_{126}^{44} (\frac{1}{6},2,3)$& $M_{PS}$ \\ 
    &&&$\Phi_{126}^{45} (\frac{-1}{6},2,\overline{3})$& $M_{PS}$ \\ 
     &&&$\Phi_{126}^{46} (\frac{-7}{6},2,\overline{3})$& $M_{PS}$ \\ 
      &&&$\Phi_{126}^{47} (\frac{-1}{2},2,8)$& $M_{PS}$ \\ 
       &&&$\Phi_{126}^{48} (\frac{1}{2},2,8)$& $M_{PS}$ \\ \cline{2-5}
& \multirow{4}{*}{$\Phi_{54} $ }  & $\Phi_{54}^1 (3,3,1)$  &   & $M_{SO(10)}$ \\  
&  & *$\Phi_{54}^2 (2,2,6)$  &  & $M_{SO(10)}$ \\ 
&  & $\Phi_{54}^3 (1,1,20')$   &   & $M_{SO(10)}$ \\ 
& & $\Phi_{54}^4 (1,1,1)$    &   & $M_{SO(10)}$ \\
\cline{2-5}
 & *$\Phi_{\overline{16}}$ &  &  &   $M_{E_6}$\\ \cline{2-5}
 &  $\Phi_{10}$ & &  & $M_{E_6}$ \\  \cline{2-5}
 & *$\Phi_{1}$ &   &  &  $M_{E_6}$ \\
\end{tabular}
\caption{Higgs masses according to the extended survival hypothesis. $M_{SM}$ denotes the standard model scale and $M_{PS}$ the Pati-Salam scale. Because of $D$ parity the Higgs in the $(3,1,10)$ have the same mass as the Higgs in the $(1,3,\overline{10})$, although they do not develop a vev. The $(2,2,15)$ isn't superheavy because a small induced vev for this representation is needed in realistic models \cite{Babu:2015bna}. In addition, we assume that only one Higgs doublet remains light. The fields marked with an asterisk correspond to Goldstone bosons.}
\label{tab:higgsmasses}
\end{table}

\noindent For the numerical integration of Eq. \eqref{eq:twolooprge} we need the boundary conditions \cite{Agashe:2014kda}
\begin{align}
\omega_{1Y}(M_Z)& = 59.0116 \\
\omega_{2L}(M_Z)& = 29.5874\\
\omega_{3C}(M_Z)& = 8.4388 \\
M_Z &= 91.1876 \text{ \ GeV}.
\end{align}
In addition, we need the two-loop matching conditions for the case that a group $G$ breaks into several factors $ G \rightarrow \prod_i G_i$ \cite{Hall:1980kf}
\begin{equation}
\omega_G - \frac{C_G}{12 \pi} = \omega_{G_i} - \frac{C_{G_i}}{12 \pi},
\end{equation}
where $C_G$ and $C_{G_i}$ are the quadratic Casimir invariants of $G$ and $G_i$, respectively. These only hold if the smaller group $G_i$ comes from one grand group $G$, as it is the case for $SU(4)_C \rightarrow SU(3)_C$. In contrast, the group $U(1)_Y$ comes from $SU(4)_C$ \textbf{and} $SU(2)_R$ and the correct matching condition reads
\begin{equation}
\omega_{1Y}= \frac{3}{5}\left( \omega_{2R}-  \frac{C_2}{12 \pi} \right) +   \frac{2}{5}\left( \omega_{4C}-  \frac{C_4}{12 \pi} \right). 
\end{equation}
\begin{figure}[t]
\centering
\includegraphics[width=0.8\linewidth]{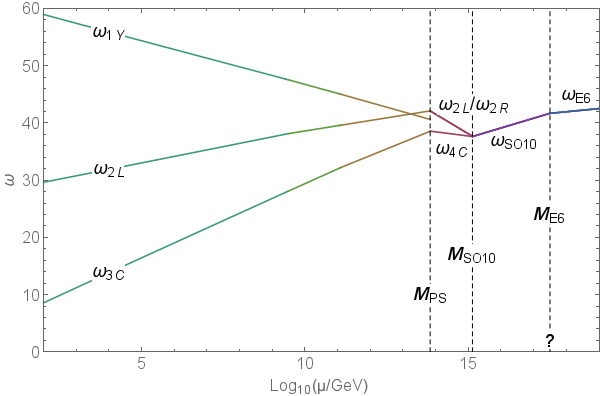}
\caption{Running of the gauge coupling without threshold corrections. The $E_6$ scale can not be calculated as we have no boundary condition left. It is shown here for illustration purposes only.}
\label{fig:2LoopWOThresholds}
\end{figure}
The result of the numerical integration is shown in figure \ref{fig:2LoopWOThresholds} and yields for the Pati-Salam scale $M_{PS} \simeq 6.8 \cdot 10^{13} $ GeV and for the $SO(10)$ scale $M_{SO(10)} \simeq 1.34 \cdot 10^{15} $ GeV. This is almost the same result as in models without exotic fermions \cite{Deshpande:1992au}. This means the exotic fermions alone are not enough to yield a proton lifetime above the present bound from Super-Kamiokande $\tau_P \gtrsim  1.6 \cdot 10^{34}$ yrs for the dominant decay mode $p \rightarrow \pi^0 e^+ $ in non-supersymmetric GUTs \cite{Miura:2016krn}.
However it is well known that threshold correction can alter these results significantly \cite{Dixit:1989ff} and we estimate the magnitude of these effects in the next section.

\subsubsection{Threshold corrections}
\label{sec:thresholdeffects}
\begin{figure}[t]
\centering
\begin{subfigure}{.5\textwidth}
  \centering
  \includegraphics[width=0.9\linewidth]{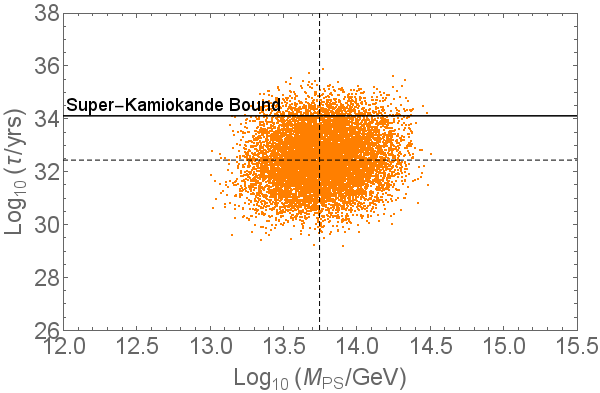}
  \caption{$R \in [\frac{1}{10}, 2 ]$}
  \label{fig:sub1}
\end{subfigure}%
\begin{subfigure}{.5\textwidth}
  \centering
  \includegraphics[width=0.9\linewidth]{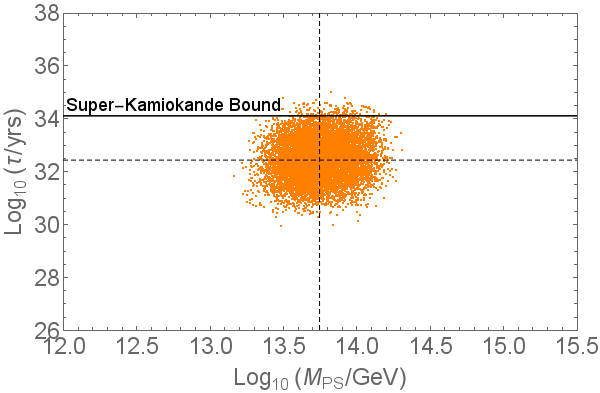}
  \caption{$R \in [ \frac{1}{5}, 2 ]$}
  \label{fig:sub2}
\end{subfigure}
\caption{Proton lifetime $\tau_P \approx \frac{M_{SO10}^4 }{m_p^5 \alpha_{SO10}^2}$, where $\alpha_{SO10}$ denotes the unified coupling constant and $m_p$ the proton mass, as a function of the Pati-Salam scale $M_{PS}$ for randomized Higgs masses $M_S = R M_V$ and two different ranges $R$. The solid black line denotes the present bound from Super-Kamiokande and the dashed lines the result when all scalars have exactly the same mass. Almost the complete possible parameter space for both ranges ranges will be explored by Super-Kamiokande and the next generation proton decay experiments in the near future \cite{Abe:2011ts}.}
\label{fig:protonscan}
\end{figure}

\begin{figure}[t]
\centering
\includegraphics[width=0.8\linewidth]{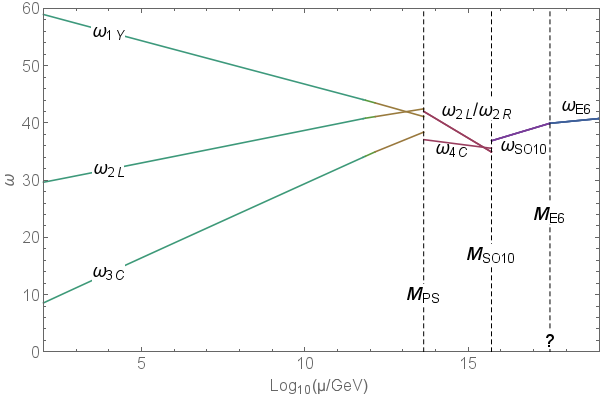}
\caption{Running of the gauge couplings with threshold corrections. The symmetry breaking scales are no longer where the couplings meet, but can lie above or below these points. The $E_6$ scale can not be calculated as we have no boundary condition left. It is shown here for illustration purposes only.}
\label{fig:withthresholds}
\end{figure}

It is unlikely that all scalars masses at a given symmetry breaking scale are all exactly degenerate and thus, for a large number of scalars, there are possibly large threshold corrections that can change the proton lifetime significantly. These threshold corrections can be written in terms of modified matching conditions \cite{Hall:1980kf}
\begin{equation}
\label{eq:thresholddef}
\omega_i(\mu)=\omega_G(\mu)-\dfrac{\lambda_i(\mu)}{12 \pi} ,
\end{equation}
where
\begin{eqnarray}
\lambda_i(\mu)= \underbrace{\left( C_G-C_i \right)}_{\lambda_i^G}  +\underbrace{Tr \left(t_{iS}^2 P_{GB} \ln \dfrac{M_S}{\mu} \right)}_{\lambda_i^S}.
\end{eqnarray}
$S$ denotes the scalar particles that are integrated out at the matching scale $\mu$,  $t_{iS}$ are the generators of $G_i$ for the heavy scalar representations and we omitted possible additional contributions from spin $1$ and spin $\frac{1}{2}$ particles. $P_{GB}$ is an operator that projects out the Goldstone bosons. The traces of the quadratic generators are often called Dynkin indices and can be found, for example, in ref. \cite{Slansky:1981yr}. We define $\eta_j^a=\ln(\frac{M_j}{M_a})$, where $j$ denotes a Higgs multiplet and $a$ is either $i$ or $u$, which denote the intermediate and unification scale, respectively. For our model we have
\begin{eqnarray}
\lambda_{3C}^{iS} &=& 3 \eta_{\Phi_{126}^{22}}^i + 15\eta_{\Phi_{126}^{23}}^i + \eta_{\Phi_{126}^{34}}^i+ \eta_{\Phi_{126}^{35}}^i  + 5\eta_{\Phi_{126}^{37}}^i +5 \eta_{\Phi_{126}^{38}}^i + 5 \eta_{\Phi_{126}^{39}}^i    +2 \eta_{\Phi_{126}^{43}}^i + 2\eta_{\Phi_{126}^{44}}^i+ 2 \eta_{\Phi_{126}^{45}}^i  \nonumber\\ && + 2 \eta_{\Phi_{126}^{46}}^i + 12 \eta_{\Phi_{126}^{47}}^i + 12\eta_{\Phi_{126}^{48}}^i \nonumber \\ [0.5em]
\lambda_{2L}^{iS}&=& \eta_{\varphi_{10}^{22}}^i + 4\eta_{\Phi_{126}^{21}}^i+12 \eta_{\Phi_{126}^{22}}^i +24 \eta_{\Phi_{126}^{23}}^i+\eta_{\Phi_{126}^{41}}^i +\eta_{\Phi_{126}^{42}}^i +3 \eta_{\Phi_{126}^{43}}^i + 3\eta_{\Phi_{126}^{44}}^i + 3\eta_{\Phi_{126}^{45}}^i  + 3\eta_{\Phi_{126}^{46}}^i \nonumber \\&& +8\eta_{\Phi_{126}^{47}}^i +8\eta_{\Phi_{126}^{48}}^i \nonumber \\ [0.5em]
\lambda_{1Y}^{iS} &=&\frac{1}{5} \left( 3 \eta_{\varphi_{10}^{22}}^i + 18\eta_{\Phi_{126}^{21}}^i +6\eta_{\Phi_{126}^{22}}^i +12\eta_{\Phi_{126}^{23}}^i +24\eta_{\Phi_{126}^{33}}^i +32\eta_{\Phi_{126}^{34}}^i+2\eta_{\Phi_{126}^{35}}^i   \right. \nonumber \\  && \hspace{.5cm} \left. +64\eta_{\Phi_{126}^{37}}^i +4\eta_{\Phi_{126}^{38}}^i+16\eta_{\Phi_{126}^{39}}^i +3\eta_{\Phi_{126}^{41}}^i + 3\eta_{\Phi_{126}^{42}}^i + 49\eta_{\Phi_{126}^{43}}^i+  \eta_{\Phi_{126}^{44}}^i  + \eta_{\Phi_{126}^{45}}^i + 49\eta_{\Phi_{126}^{46}}^i \right. \nonumber \\ && \hspace{.5cm} \left.+24\eta_{\Phi_{126}^{47}}^i +24\eta_{\Phi_{126}^{48}}^i \right) \nonumber \\
\end{eqnarray}
and
\begin{equation}
\lambda_{4C}^{uS} = 2 \eta_{\varphi_{10}^1}^u + 2 \eta_{\Phi_{126}^1}^u+ 16\eta_{\Phi_{54}^3} \quad
\lambda_{2L}^{uS} = 12 \eta_{\Phi_{54}^1}^u \quad
\lambda_{2R}^{uS} = 12 \eta_{\Phi_{54}^1}^u .
\end{equation}
To approximate the threshold corrections we choose the scalar masses randomly in a given range $M_S = R M_V$, where previous studies used, for example, $R \in [ \frac{1}{10}, 10]$ \cite{Mohapatra:1992dx,Parida:1987bp} or more conservative $R \in [ \frac{1}{10}, 2]$ \cite{Babu:2015bna}. With a given randomized set of scalar masses it is possible to compute $\lambda_i^S$, which then can be used to compute the updated scales $M_{PS}$, $M_{SO(10)}$ and the updated proton lifetime $\tau_P$. For a fixed set of randomized Higgs masses the process is iterated until convergence is reached. We have already seen that the effect of the exotic fermions on the RGE running is small and therefore the results of the parameter scan do not depend significantly on the masses of the exotics. The following results are for fixed masses of the exotic fermions as given in Eq. \eqref{eq:specificscenario}. 
We only consider gauge mediated proton decay and the corresponding partial decay width for the dominant decay channel is given by~\cite{Nath:2006ut}
\begin{align}\label{Gamma}
&\Gamma(p\to \pi^0 e^+)=\frac{\pi\, m_p\,\alpha _G^2 }{4
   f_{\pi}^2}  |\alpha|^2 A_L^2 (D+F+1)^2 \\
  \nonumber &\times\left(A_{SR}^2 \left(\frac{1}{M_{(X',Y')}^2}+\frac{1}{M_{(X,Y)}^2}\right)^2 + \frac{4 A_{SL}^2}{M_{(X,Y)}^4}\right)\,,
\end{align}
where $m_p$ is the proton mass, $\alpha_G$ is the unified gauge coupling at the GUT scale and $f_\pi=139\,\mathrm{MeV}$, $\alpha=0.009 \mathrm{ \ GeV^3}$ and $D+F=1.267$ are phenomenological factors obtained in the chiral perturbation theory and lattice studies. $M_{(X,Y)}$ and $M_{(X',Y')}$ denote the masses of the gauge bosons with standard model quantum numbers $(3,2,-5/6)$ and $(3,2,+1/6)$. The renormalization group running of the effective proton decay operator from the proton mass to the electroweak scale is taken into account by $A_L\approx 1.4$ and the running from the GUT scale to the electroweak scale by $A_{SR}$ and $A_{SL}$ \cite{Munoz:1986kq}:
\begin{equation}
A_{SL(R)} = \prod_{i=1}^n \prod_{s}^{M_Z \leq m_{s} < M_U} \left[ \dfrac{\alpha_i(m_{s+1})}{\alpha_i(m_{s})}\right]^{\dfrac{\gamma_{L(R)(s)i}}{a_i(m_{s+1}-m_{s})}}\nonumber
\end{equation}
where 
\begin{equation}
\gamma_{L (M_Z)} = \gamma_{L (M_{D^1})} = \gamma_{L (M_{L^1})} =  \left\lbrace \dfrac{23}{20},\dfrac{9}{4},2\right\rbrace ; \hspace{.5cm}
\gamma_{R (M_Z)} = \gamma_{R (M_{D^1})} = \gamma_{R (M_{L^1})} = \left\lbrace \dfrac{11}{20},\dfrac{9}{4},2 \right\rbrace ; \end{equation}
and
\begin{equation}
\gamma_{L/R (M_{PS})} = \left\lbrace \dfrac{15}{4},\dfrac{9}{4},\dfrac{9}{4}\right\rbrace. 
\end{equation}
Here the $a_i$'s are again the one-loop beta-function coefficients as given in Eq. \eqref{eq:smbetacoefficients} - \eqref{eq:psbetacoefficients} and the relevant scales ($s=1,2,3,4,5$) are $M_U, M_{PS}, M_{D^1} , M_{L^1} , M_Z$. 
The proton lifetime and Pati-Salam mass scale range, resulting from randomized Higgs masses, are shown in Figure \ref{fig:protonscan}. We can see that even for a very conservative range $R \in [ \frac{1}{5}, 2]$ several configurations are in agreement with the latest Super-Kamiokande data. This confirms the findings of a recent study \cite{Babu:2015bna} and shows that the proton lifetime can be long enough in our model.
The maximal $SO(10)$ scale for this range is $M_{SO(10)}^\text{max} \approx 7.7 \cdot 10^{15}$ GeV and the maximal proton lifetime $\tau_p^\text{max} \approx 1.25 \cdot 10^{35}$ yrs. As a consequence of the threshold correction the scales are no longer where the coupling constants meet at a point, but lie above or below these points. An example plot for the coupling strengths for a randomized set of Higgs masses is shown in Figure \ref{fig:withthresholds}.

\section{Production}

The exotic neutrino $N_E$ can only be a realistic dark matter candidate if it can be produced with a correct relic density. As discussed above, we expect that even the lightest generation of the exotic neutrinos is superheavy. At a first glance this could be problematic, because there is an upper bound on the mass of dark matter candidates $m_{DM}<340$ TeV from unitarity \cite{PhysRevLett.64.615}.

However, this bound only holds for dark matter particles that were in thermal equilibrium in the early universe and it was shown in Refs. \cite{Chung:1998bt,Kolb:1998ki,Chung:1998zb} that superheavy dark matter can be produced non-thermally with a correct relic abundance. 

In principle, superheavy dark matter can be produced through scattering
during the reheating process or gravitationally at the end of inflation \cite{Chung:1998bt}. However, for dark matter masses below $10^{11}$ GeV the non-thermal production from the thermal bath dominates \cite{Feldstein:2013uha}.

An interesting aspect of this production mechanism is that it could be used to probe the reheating temperature $T_{RH}$, because demanding a correct dark matter relic density yields $30 \lesssim \frac{M_{DM}}{T_{RH}}\lesssim 10^{3.5} $ \cite{Feldstein:2013uha}. 
 
\section{Conclusions and Discussion}

In summary, we have shown that $E_6$ unification incorporates an inherent, viable dark matter candidate that could be detected in the near future. We have argued that the lightest exotic fermion can be stable through a remnant $\mathbb{Z}_2$ symmetry and discussed which breaking chains are possible with Higgs representations that couple to the fermions and leave the $\mathbb{Z}_2$ symmetry unbroken. Moreover, we have computed the consequences of the various breaking chains on the Yukawa sector. With these information at hand, we have discussed the viability of all exotic fermions as dark matter candidates and argued that only the neutral component $N_E$ of the exotic lepton doublet is a viable, interesting candidate. Then we have presented a scenario where $N_E$ is the lightest exotic fermion and therefore stable. The masses of the exotic fermions were estimated by using the fit results of a recent study for the Yukawa couplings and we have found ${6 \cdot 10^8 \text{ \ GeV} \lesssim m_{N_E} \lesssim 5 \cdot 10^{12} \text{ \ GeV} }$. Moreover, we calculated the RGEs for the gauge couplings in the presence of one generation of the exotic fermions and checked that the proton lifetime can be long enough through threshold corrections.  
The exotic superheavy neutrino $N_E$ carries hypercharge and its cross section for scattering with Xenon is $ \sigma_{DMXe} \simeq 1.68 \cdot 10^{-31} \mathrm{ \ cm^2}$. Therefore it could be detected by the next generation of direct detection experiments, like XENON1T \cite{Aprile:2012zx} or DARWIN \cite{Baudis:2014naa}. Moreover, it can be produced non-thermally with a correct relic density and a its detection could be used to deduce the reheating temperature.

\section*{Acknowledgments}

The author wishes to thank Ulrich Nierste, Robert Ziegler, Tommy Ohlsson, Paolo Panci, Saki Khan, Rober Fegert and Manuel Masip for helpful discussions and acknowledges the support by the DFG-funded Doctoral School KSETA.

\appendix

\section{Standard Model and Subgroup Embeddings}
\label{sec:smemebdding}

The computations in this section where done using LieArt \cite{2012arXiv1206.6379F}. In order to specify the particle content of an $E_6$ representation, we must specify the embedding of $G_{SM}$ in $E_6$. This means we need to identify the standard model Cartan generators among the $E_6$ Cartan generators. It is then convenient to define the corresponding dual Cartan generators $\tilde h_i$. These act on an arbitrary weight $\mu$ in the Dynkin basis via the usual Euclidean scalar product and yield the eigenvalue of $\mu$ corresponding to the Cartan generator $H_i$
$$ \tilde h_i \cdot \mu = H_i(\mu). $$
These dual Cartan generators are often called charge axes, because acting with these on a weight in the Dynkin basis yields the corresponding quantum numbers. 

\vspace{0.3cm}

\begin{figure}[H]
\centering
\includegraphics[width=0.3\linewidth]{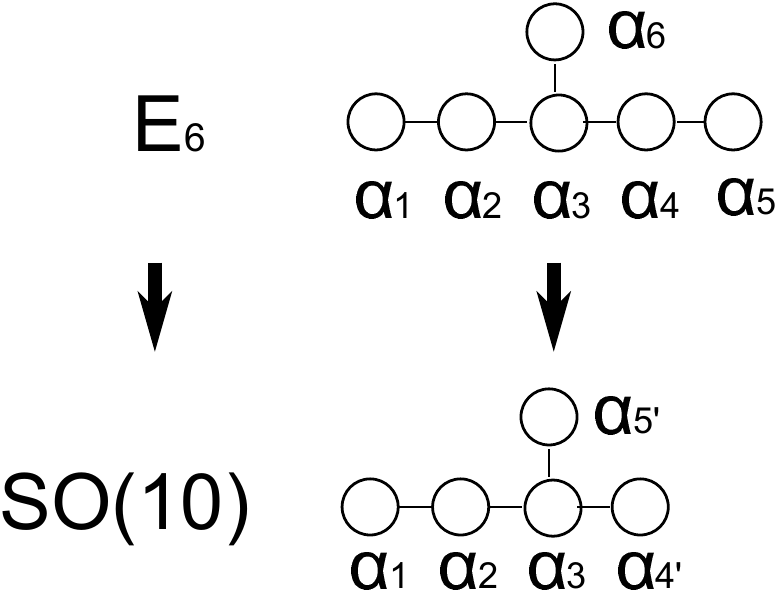}
\caption{Embedding of $SO(10)$ in $E_6$}
\label{fig:SO10inE6}
\end{figure}

We present now our embedding of $G_{SM}$ in $E_6$ that can be used in models with $SO(10)$ as intermediate symmetry. By looking at the Dynkin diagrams of $E_6$ and $SO(10)$, we can identify how $SO(10)$ can be embedded in $E_6$. One possibility is shown in Fig. \ref{fig:SO10inE6}. 

\begin{figure}[H]
\centering
\includegraphics[width=0.5\linewidth]{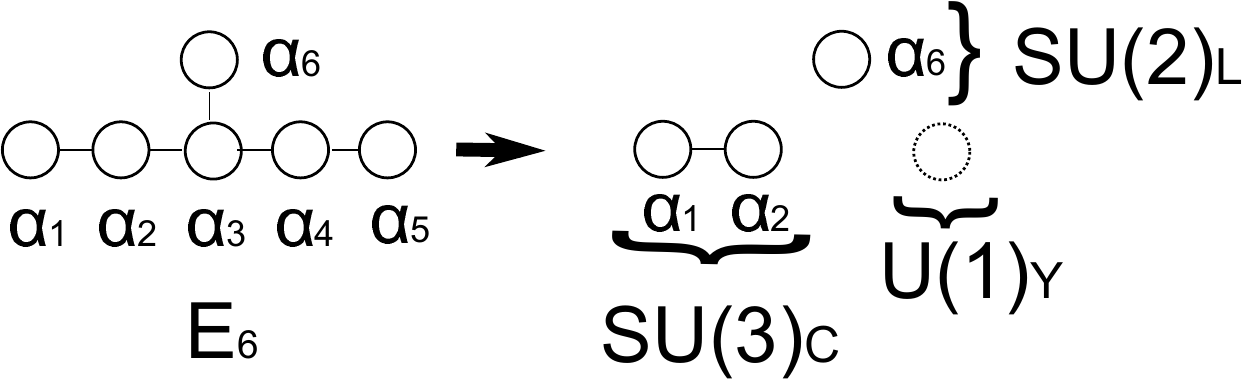}
\caption{One possible embedding of $G_{SM}$ in $E_6$.}
\label{fig:SMinE6}
\end{figure}

We need an embedding of $G_{SM}$, which does not get broken by the breaking $E_6 \rightarrow SO(10)$. In Fig. \ref{fig:SO10inE6} we can see that the first three nodes and the sixth node remain unchanged, whereas the fourth and fifth node become one. Each node corresponds to a simple root $\alpha_i$ and a Cartan generator $H_{\alpha_i}$. Therefore, the standard model Cartan generators must correspond to nodes that remain intact at the $SO(10)$ scale. One possibility to accomplish this is that in the $E_6$ diagram the first two nodes correspond to $SU(3)_C$ and the sixth node to $SU(2)_L$. (The standard model hypercharge $U(1)_Y$ is more complicated as it is a linear combination of $U(1)$ factors.) This is shown in Fig \ref{fig:SMinE6}. 

The dual standard model Cartan generators corresponding to this embedding are
\begin{align}
\label{eq:dualcartan}
\tilde h_1^{SU(3)_c} = \frac{1}{2} (1,1,0,0,0,0) \\
\tilde h_2^{SU(3)_c}=  \frac{1}{2\sqrt{3}} (1,-1,0,0,0,0) \\
\tilde h_1^{SU(2)_L} = \frac{1}{2} (0,0,0,0,0,1) \\
\tilde Y =  (-\frac{1}{3},-\frac{2}{3},-1,-1,-1,-\frac{1}{2}) \, .
\end{align}

\bibliographystyle{h-physrev}
\bibliography{bib}

\end{document}